\documentclass{aa}
\usepackage{graphicx}
\usepackage{hyperref}
\usepackage{amssymb}
\usepackage{float}
\usepackage{color}
\usepackage{epstopdf}
\usepackage{ctable}
\usepackage{multirow}

\newcommand{\amin}[1]{{#1}$^\prime$}
\newcommand{\asec}[1]{{#1}$^{\prime\prime}$}

\begin{document}

\title{Ultra-steep spectrum emission in the merging galaxy cluster Abell~1914}
\titlerunning{A1914-LOFAR}

\author{S.~Mandal\thanks{E-mail: mandal@strw.leidenuniv.nl}\inst{1}\and
H.~T.~Intema\inst{1}\and T.~W.~Shimwell\inst{1,2}\and R.~J.~van~Weeren\inst{1}\and A.~Botteon\inst{3,4}\and H.~J.~A.~R\"ottgering\inst{1}\and D.~N.~Hoang\inst{1}\and G.~Brunetti\inst{4}\and F.~de~Gasperin\inst{5} \and S.~Giacintucci\inst{9} \and H.~Hoekstra\inst{1}\and A.~Stroe \inst{6}\and M.~Br\"uggen\inst{5}\and R.~Cassano\inst{4}\and A.~Shulevski\inst{7}\and A.~Drabent\inst{8}\and D.~Rafferty\inst{5}}

\institute{Leiden Observatory, Leiden University, PO Box 9513, NL-2300 RA Leiden, The Netherlands \and
ASTRON, the Netherlands Institute for Radio Astronomy, Postbus 2, NL-7990 AA Dwingeloo, The Netherlands \and
Dipartimento di Fisica e Astronomia, Universit\`{a} di Bologna, via P.~Gobetti 93/2, I-40129 Bologna, Italy \and 
INAF - IRA, via P.~Gobetti 101, I-40129 Bologna, Italy \and
Hamburger Sternwarte, Gojenbergsweg 112, D-21029 Hamburg, Germany \and
European Southern Observatory, Karl-Schwarzschild-Str. 2, D-85748 Garching, Germany \and
Anton Pannekoek Institute for Astronomy, University of Amsterdam, Postbus 94249, 1090 GE Amsterdam, The Netherlands \and
Th\"uringer Landessternwarte, Sternwarte 5, D-07778 Tautenburg, Germany \and
Naval Research Laboratory, 4555 Overlook Avenue SW, Code 7213, Washington, DC 20375, USA
}
\authorrunning{S.~Mandal et al.}

\date{\today}
\date{Accepted XXX. Received YYY; in original form ZZZ}


\label{firstpage}

\abstract{A number of radio observations have revealed the presence of large synchrotron-emitting sources associated with the intra-cluster medium. There is strong observational evidence that the emitting particles have been (re-)accelerated by shocks and turbulence generated during merger events. The particles that are accelerated are thought to have higher initial energies than those in the thermal pool but the origin of such mildly relativistic particles remains uncertain and needs to be further investigated. The galaxy cluster Abell~1914 is a massive galaxy cluster in which X-ray observations show clear evidence of merging activity. We carried out radio observations of this cluster with the LOw Frequency ARay (LOFAR) at 150~MHz and the Giant Metrewave Radio Telescope (GMRT) at 610~MHz. We also analysed Very Large Array (VLA) 1.4~GHz data, archival GMRT 325~MHz data, CFHT weak lensing data and Chandra observations. Our analysis shows that the ultra-steep spectrum source (4C38.39; $\alpha \lesssim -2$), previously thought to be part of a radio halo, is a distinct source with properties that are consistent with revived fossil plasma sources. Finally, we detect some diffuse emission to the west of the source 4C38.39 that could belong to a radio halo.}


\maketitle

\section{Introduction}

Galaxy clusters are the largest gravitationally bound systems in the Universe. They grow through the accumulation of smaller groups of galaxies and through major mergers with other massive clusters. A huge amount of gravitational binding energy ($\sim 10^{64}$~ergs) is released when massive galaxy clusters merge (e.g. \citealt{kravstov12}). This strongly affects the physical properties of the different properties of clusters, for example the density distribution and velocity dispersion of galaxies, and the temperature, metallicity, and density distribution of the (X-ray-emitting) thermal intra-cluster medium (ICM). Direct evidence for these merging events is seen using X-ray observations, which show the disturbed surface brightness of the ICM and the presence of density and temperature jumps (shocks and cold fronts: \citealt{markevitch07}). The shocks and turbulence that are generated in the ICM might also amplify magnetic fields ($\sim\mu$G) and accelerate relativistic particles (Lorentz facotr $\gamma >> 1000$), resulting in megaparsec-scale synchrotron emission regions (for review: \citealt{feretti12}, \citealt{brunettijones14}). The spectral index $\alpha$ of the synchrotron emission is generally steep: $\alpha \lesssim -1$ with $S\propto\nu^{\alpha}$, where $\nu$ is the observed frequency and $S$ is the measured flux density. The steep spectral index suggests that the synchrotron emission is relatively bright at low radio frequencies.
\begin{figure*}
\begin{center}
\resizebox{0.95\hsize}{!}{
\includegraphics[angle=0]{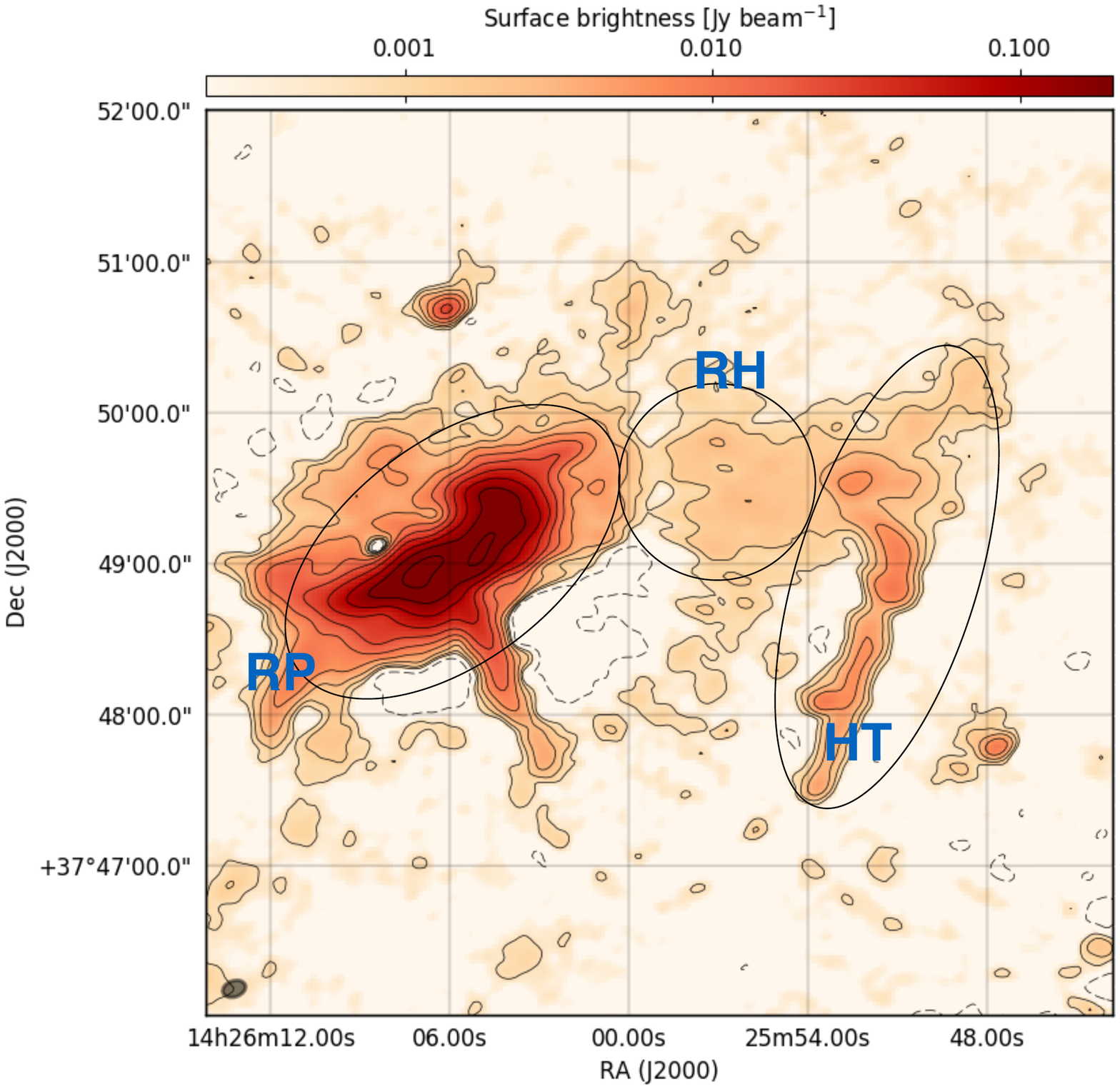}
}
\caption{Full-resolution (\asec{9} $\times$ \asec{6}) LOFAR 120-160~MHz image of Abell~1914 with regions of interest circled and labelled. The black contours and dashed lines indicate the significance of the radio emission at (1,2,4,...)$\times$3 $\times$ $\sigma_{\rm{LOFAR}}$ and -3$\times$ $\sigma_{\rm{LOFAR}}$ levels, respectively, where $\sigma_{\rm{LOFAR}}=150$~$\mu$Jy\,beam$^{-1}$.} 
\label{fig:lofar150}
\end{center}
\end{figure*}

\begin{figure*}
\begin{center}
\resizebox{0.95\hsize}{!}{
\includegraphics[angle=0]{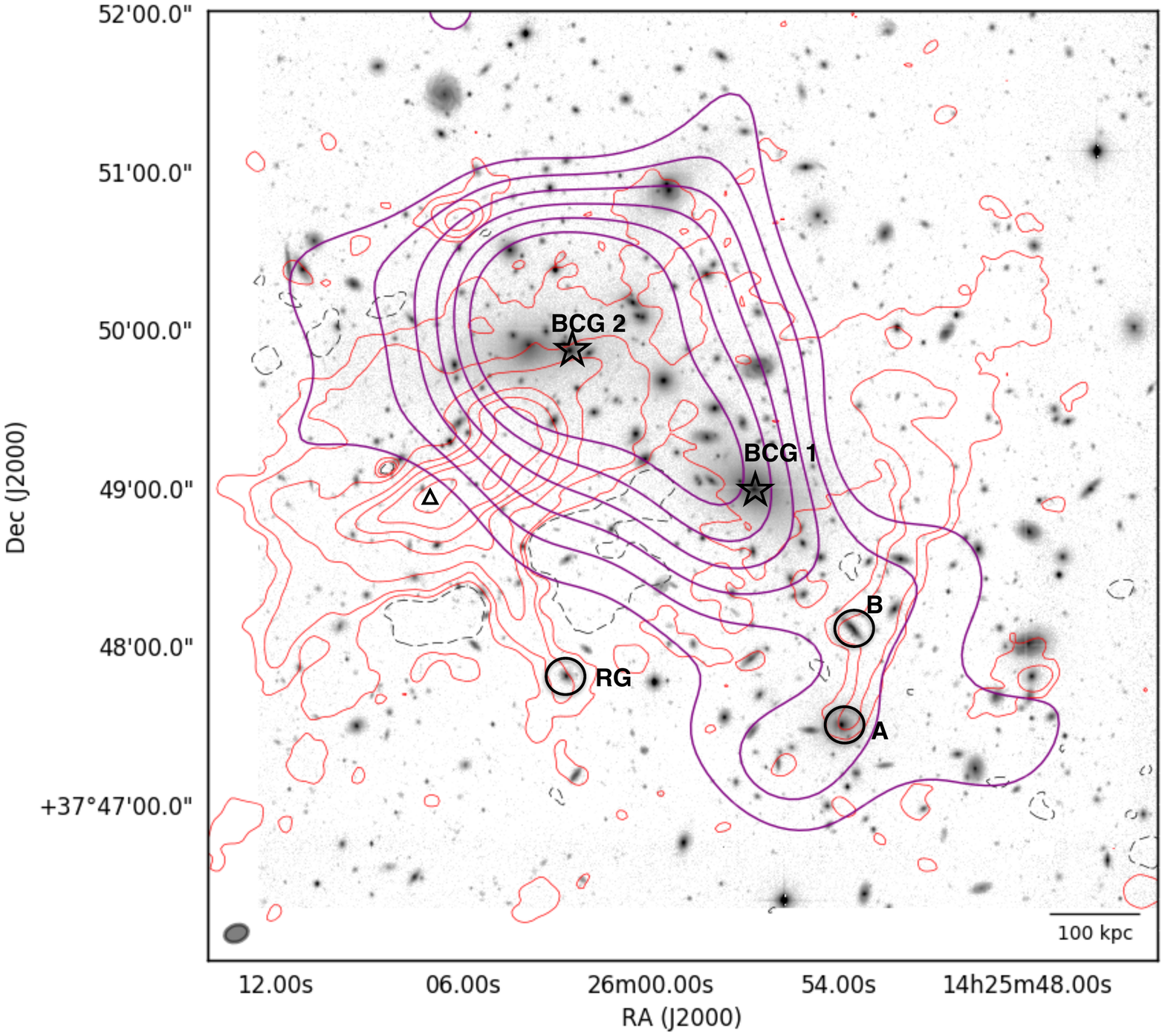}
}
\caption{Full-resolution (\asec{9} $\times$ \asec{6}) LOFAR image overlaid on a CFHT $r$-band image. The LOFAR contours (solid red) show the (1,2,4,...)$\times$3 $\times$ $\sigma_{\rm{LOFAR}}$ levels and the dashed red contours show the -3$\sigma_{\rm{LOFAR}}$ contour where $\sigma_{\rm{LOFAR}}=150$~$\mu$Jy\,beam$^{-1}$. The purple contours represent the weak lensing measurement (CFHT data: section 2.5) showing the dark matter distribution. The two stars indicate the position of the two brightest cluster galaxies (BCG) in the cluster. The triangle indicates the position of the peak radio emission. The circle A indicates the location of the galaxy that is the optical counterpart of source HT. Circle B indicates a radio-loud galaxy that is not associated with source HT. Circle RG indicates the location of the galaxy that could be a possible counterpart of RP (discussed in section 4).}
\label{fig:sdss-wl-radio}
\end{center}
\end{figure*}

While the nomenclature used to describe the emission needs refinement, these synchrotron sources are most commonly classified as either radio haloes or radio relics (\citealt{feretti12}). There are about 50 radio haloes that have been discovered; these are extended ($\gtrsim 1$~Mpc) diffuse radio sources at the centre of merging clusters (e.g. \citealt{cassano10a}; \citealt{cuciti15}; \citealt{kale15}) and have a morphology similar to the X-ray morphology.

One of the theoretical models to explain their formation is merger-driven turbulent re-acceleration (\citealt{brunetti01}, \citealt{petrosian01} \citealt{brunettilazarian07}, \citealt{pinzke17}, \citealt{brunettizimmer17}). Although alternative models have been proposed, i.e. secondary models, in which the synchrotron emitting electrons are continuously injected by inelastic collisions between cosmic ray protons and thermal protons (e.g. \citealt{dennison80}; \citealt{ensslin11}). The process of generation of secondary particles via proton-proton collisions is thought to play a minor role due to current upper limits in gamma ray observations (\citealt{ackermann14}. \citealt{ackermann16}, \citealt{brunetti12}, \citealt{brunettizimmer17}, \citealt{zandanel14}) unless this mechanism is combined with turbulent re-acceleration. Radio relics are usually found in cluster outskirts and are thought to trace merger-induced shock waves (e.g. \citealt{ensslin98}; \citealt{vanweeren10}; \citealt{vanweeren12}; \citealt{roettiger99}). A connection between relics and shocks has been established by X-ray detections of shocks at the relic location (\citealt{finogenov10}, \citealt{akamatsu13}, \citealt{shimwell15}, \citealt{botteon16}, \citealt{eckert16a}). However, in some cases, the luminosity of radio relics is much higher than expected from the acceleration efficiency of thermal electrons from diffuse shock acceleration (DSA; e.g. \citealt{brunettijones14}, \citealt{botteon16}, \citealt{vanweeren16a} etc.). Shocks are also able to re-accelerate these electrons via Fermi-type processes. Simulations indicate that such re-acceleration of fossil relativistic electrons should be more efficient than the acceleration of electrons from the thermal pool of the ICM (e.g. \citealt{kangryu15}). Relics are elongated regions with sizes up to 1-2~Mpc long, which have a convex morphology with respect to the cluster centre, exhibit a radio spectral index steepening towards the cluster centre, and are linearly polarised (>10\%-30\%  at GHz frequencies; \citealt{vanweeren10} \citealt{bonafede12}, \citealt{stroe13}, \citealt{gasperin15}, \citealt{kierdorf17}).
Radio phoenixes are another, less widely studied class of diffuse radio sources in the ICM. These are thought to be a manifestation of fossil plasma in galaxy clusters (e.g. \citealt{slee01}; \citealt{ensslinbruggen02}; \citealt{kempner04}; \citealt{vanweeren09}; \citealt{gasperin15}). Likely candidates for the fossil electrons are old lobes of radio galaxies that have ultra-steep spectra (USS) owing to synchrotron and inverse Compton (IC) losses. When a merger shock passes through an old lobe (or radio tail), it compresses the fossil radio plasma, thereby re-energizing the electrons and boosting their visibility \citealt{eg01}. As the sound speed in the fossil plasma is much larger than the surrounding ICM and the shocks have a low Mach number ($\mathcal{M}<2$), this may result in subsonic compression waves generated within the plasma (\citealt{eg01},  \citealt{ensslinbruggen02}). Sources powered by this mechanism can maintain electrons at higher energies than what radiative cooling would allow. Although deposited locally, fossil plasma can, over time, occupy a significant volume of the ICM due to turbulent diffusion. The relative importance of these proposed mechanisms to explain the plethora of diffuse cluster radio sources is still unclear.

\begin{figure*}
\begin{center}
\resizebox{0.49\hsize}{!}{
\includegraphics[angle=0]{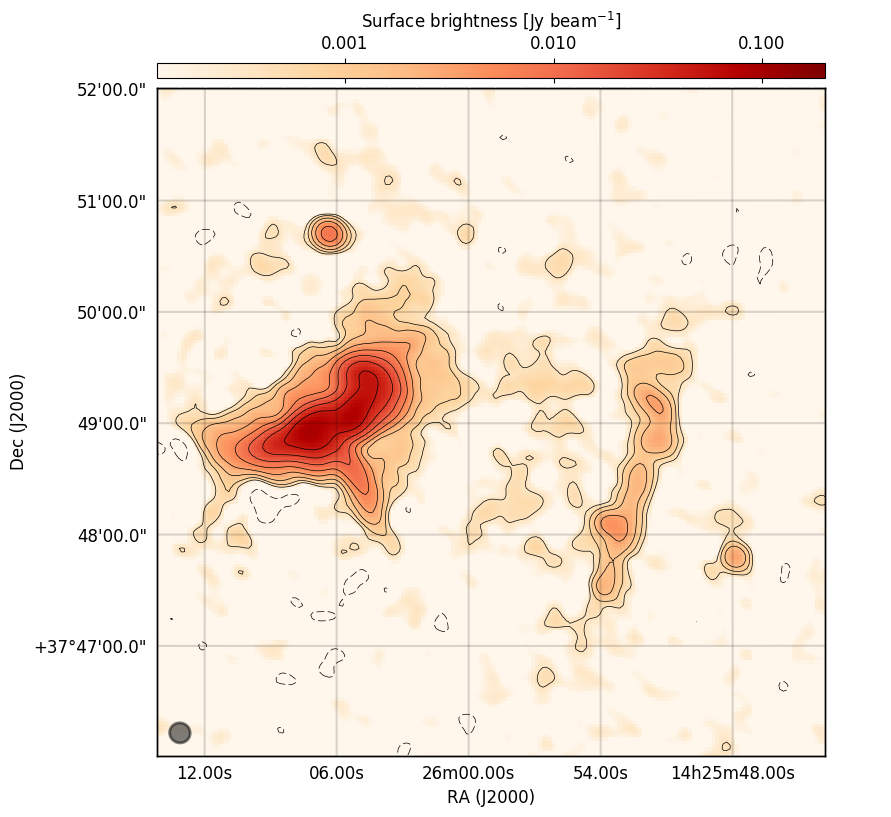}
}
\resizebox{0.49\hsize}{!}{
\includegraphics[angle=0]{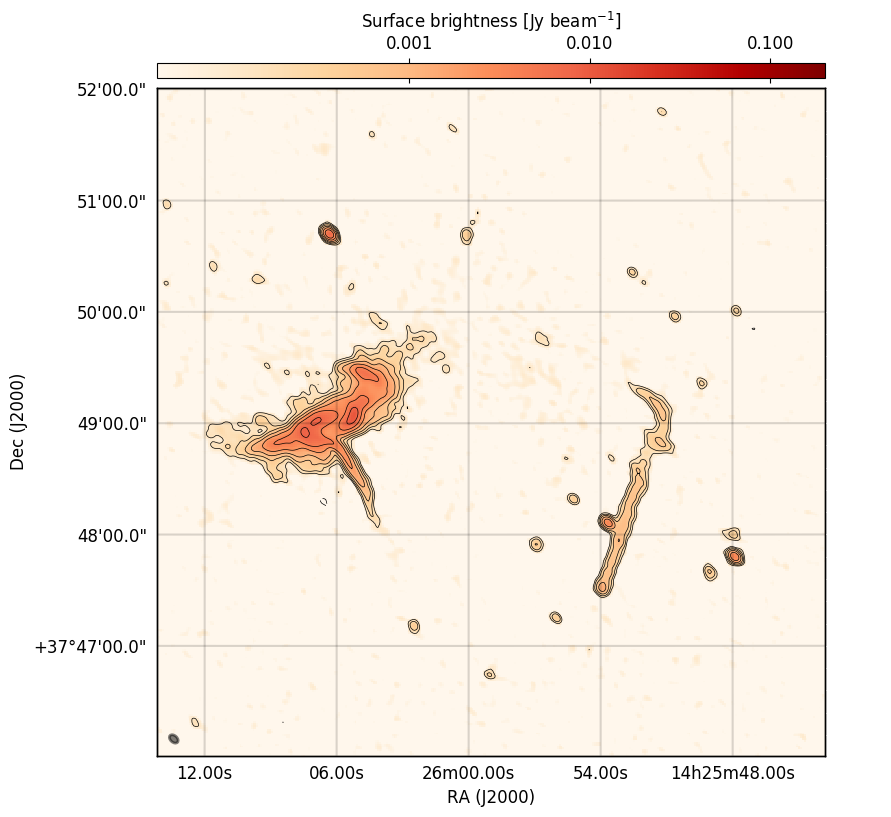}
}
\caption{Full-resolution GMRT images of Abell~1914 at 325~MHz (\textit{left}: \asec{11} $\times$ \asec{10}) and 610~MHz (\textit{right}: \asec{5} $\times$ \asec{4}). The black contours and dashed lines indicate the significance of the radio emission at (1,2,4,...)$\times$3 $\times$ $\sigma_{\rm{GMRT,325/610}}$ and -3 $\times$ $\sigma_{\rm{GMRT,325/610}}$ levels, respectively, where $\sigma_{\rm{GMRT,325}} = 130$~$\mu$Jy\,beam$^{-1}$ and $\sigma_{\rm{GMRT,610}} = 40$~$\mu$Jy\,beam$^{-1}$.}
\label{fig:gmrt-high}
\end{center}
\end{figure*}

\begin{figure*}
\begin{center}
\resizebox{0.49\hsize}{!}{
\includegraphics[angle=0]{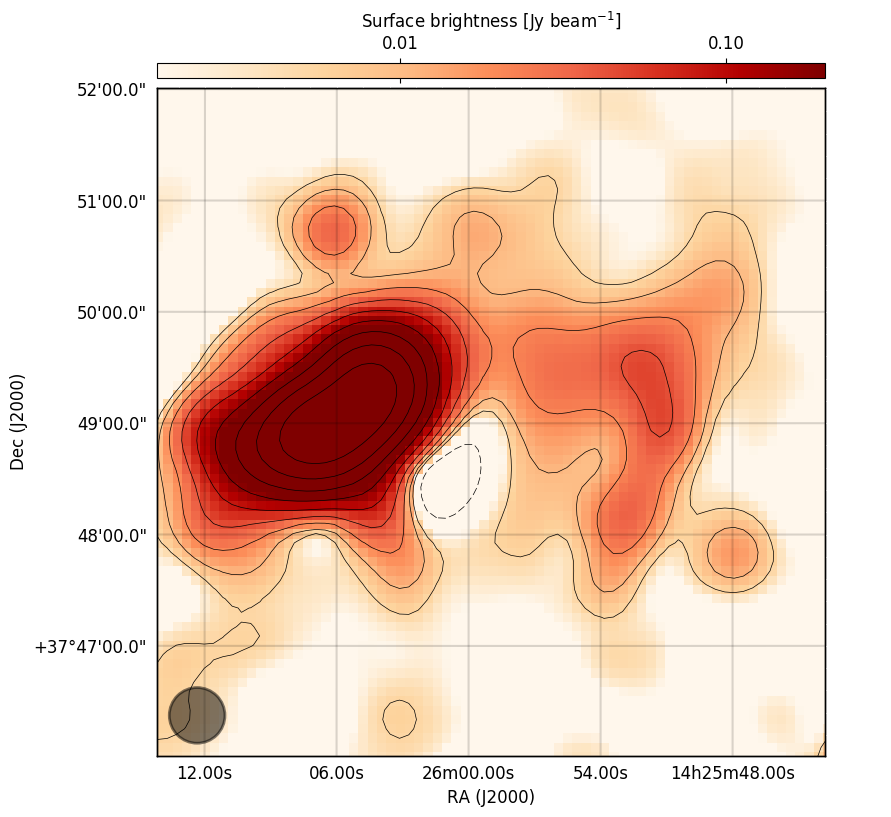}
}
\resizebox{0.49\hsize}{!}{
\includegraphics[angle=0]{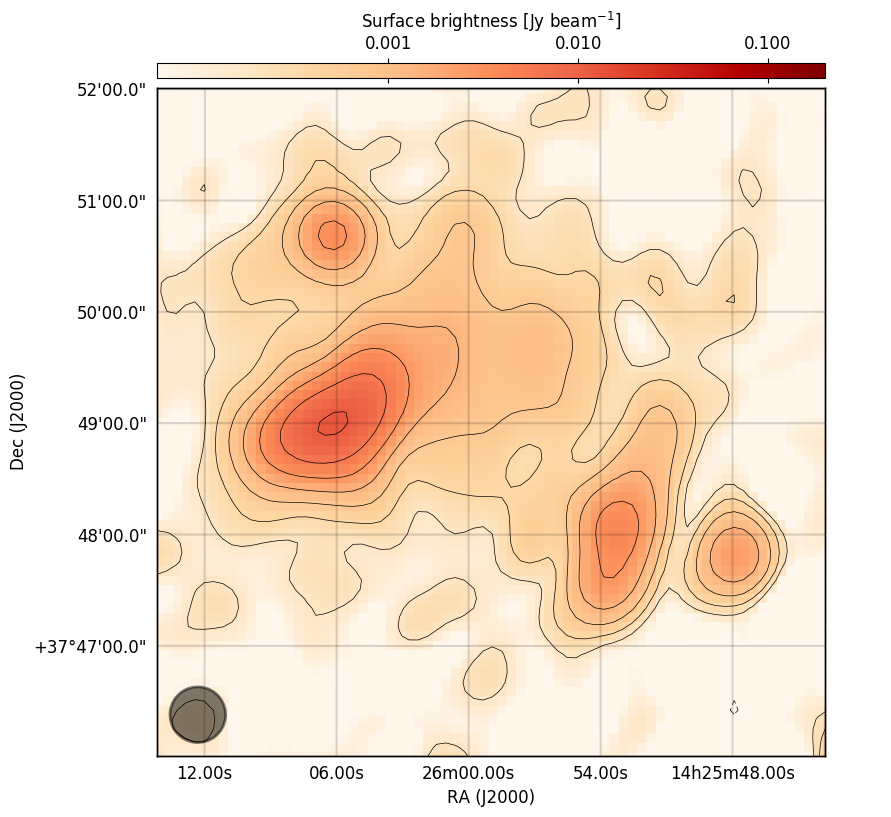}
}
\caption{Left: 150~MHz LOFAR image (\asec{30} $\times$ \asec{30}) in which the contours show the (1,2,4,...)$\times$3 $\times$ $\sigma_{\rm{LOFAR,150}}$ levels, where $\sigma_{\rm{LOFAR,150}} = 1.8$~$m$Jy\,beam$^{-1}$. Right: 1.4~GHz VLA image (\asec{26} $\times$ \asec{14}) where the contours show the (1,2,4,...)$\times$3 $\times$ $\sigma_{\rm{VLA,1400}}$ levels, where $\sigma_{\rm{VLA,1400}} = 40$~$\mu$Jy\,beam$^{-1}$.}
\label{fig:lofar-vla}
\end{center}
\end{figure*}

In this paper we study extended radio emission in the galaxy cluster Abell~1914, a merging cluster at z=0.168 with an asymmetric X-ray brightness distribution (\citealt{govoni04}; G04 hereafter), a high velocity dispersion of $\sigma_V = 1210_{-110}^{+125} \rm{km s}^{-1}$ \citep{barrena13}, and a  mass of $M_{500} = 7.3 \pm0.3 \times 10^{14} h^{-1} M_{\odot}$ \citep{planck}. The cluster has a luminosity $L_X$=(0.1-2.4 keV)=17.93 $\times 10^{44}$ $h_{50}^{-2}$erg $\rm{s}^{-1}$ (\citealt{ebeling96}) and temperature $kT_X \sim$ 9 keV (\citealt{baldi07}; \citealt{maughan08}). In the morphological and dynamical analysis performed by \cite{buote96}, the values of the power ratios $P1/P0$ and $P3/P0$ for this cluster were found to be representative of a disturbed system. Furthermore, \cite{botteon18} found surface brightness and temperature jumps in the ICM. At radio wavelengths, the cluster is known to host an extended, very bright, steep spectrum, and unpolarised radio source (\citealt{bacchi03}; B03 hereafter), namely 4C38.39 (\citealt{roland85}; \citealt{kg94}; \citealt{kempnersarazin01}). The total flux density as reported in B03 is $20 \pm 3$~mJy at 1.4~GHz, while the spectral index is found to be $\alpha \sim -1.8$. At the resolution of radio data used in the study by B03 (Figure~6) it was suggested that 4C38.39 is a part of a radio halo. However, as we discuss in more detail, the bulk of the radio emission does not align well with the X-ray emission of the ICM nor with the galaxy density distribution. Higher resolution radio observations are essential for shedding more light on the true nature of this source.

In this paper we study the morphology and spectral properties of the radio emission in Abell~1914 at higher angular resolution and better sensitivity than previously using new LOw Frequency ARray (LOFAR) 150~MHz and Giant Metrewave Radio Telescope (GMRT) 610~MHz observations, combined with archival Very Large Array (VLA) 1.4~GHz and GMRT 325~MHz observations, and complemented with optical Canada France Hawaii Telescope (CFHT) and \textit{Chandra} X-ray data. The radio telescope LOFAR \citep{vanhaarlem13} observes in the relatively poorly explored frequency range between  10 MHz and 90~MHz using the Low Band Antenna (LBA) and 110-240~MHz using the High Band Antenna (HBA). The high sensitivity and abundance of short and long baselines makes LOFAR an ideal instrument to study low-surface brightness diffuse steep-spectrum objects in clusters (e.g. \citealt{shimwell16}; \citealt{vanweeren16a}; \citealt{hoang17}; \citealt{wilber18}; \citealt{botteon18b}). 
 
This paper is structured as follows: We present the observations and data processing in Section 2. In Section 3 we show the observational results followed by a discussion of these results in Section 4. In this paper, we assume a $\Lambda$CDM cosmology with $\Omega_m$ = 0.3, $\Omega_\Lambda$ = 0.7 and $H_0$=70 km $\rm{s}^{-1}$~Mpc$^{-1}$. At the redshift of the Abell~1914 ($z = 0.168$) the luminosity distance is 808.5~Mpc and 1~arcsec corresponds to 2.873~kpc. All the coordinates are given in epoch J2000.
 
\section{Data reduction}

\ctable[botcap,center,star,
caption = {Details of the radio observations towards Abell~1914.},
label = tab:obs
]{l c c c c}{
}{
\FL {} & LOFAR 150~MHz & GMRT 325~MHz & GMRT 610~MHz & VLA 1.4~GHz
\ML Pointing RA,Dec & 14:26:01,+37:49:38 & 14:26:03,+37:49:32 & 14:26:02,+37:49:38 & 14:26:01,+37:49:38
\NN Configuration & \texttt{HBA\_DUAL\_INNER} & N/A & N/A & C and D
\NN Observation date & April 13, 2013 & August 29, 2009 & Dec. 4, 2016 & June 28, 2000 \& Sept 25, 2000
\NN On-source time (hr) & 8.0 & 6 & 6 & 5
\NN Freq. coverage (MHz) & 120-168 & 308-340 & 594-626 & 1.365-1.435
\LL}

\ctable[botcap,center,star,
caption = {Imaging parameters and image properties of the LOFAR, GMRT, and VLA maps of Abell~1914.},
label = tab:image
]{l c c c c c c c}{
\tnote[a]{The beam position angle (PA) is measured in degrees from north through east.}
}{
\FL Data &  Frequency & robust & taper & $\Theta_{FWHM}$ & $\sigma_{rms}$ & Figure
\NN     & (MHz) &  & (\asec{})  & \asec{} $\times$ \asec{}, (PA\tmark[a]) & $\mu$Jy beam$^{-1}$ \ML LOFAR & 150 & +0.5 & 0 & 9$\times$6, (-77.0) & 150 & \ref{fig:lofar150}
\NN LOFAR & 150 & 0.0 & 22 & 30$\times$30, (0.0) & 1500 & \ref{fig:lofar-vla}
\NN GMRT  & 325 & 0.0 & 0 & 12$\times$11, (68.0) & 100 & \ref{fig:gmrt-high}
\NN GMRT  & 610 & 0.0 & 0 & 6$\times$5, (47.0) & 55 & \ref{fig:gmrt-high}  
\NN VLA   & 1400 & 0.0 & 22 & 30$\times$30, (0.0)& 51 & \ref{fig:lofar-vla}
\LL}


\subsection{LOFAR data reduction}

We observed Abell~1914 in \texttt{HBA\_DUAL\_INNER} mode on 13 April 2013. The observation (with ID: 248837) was centred on RA 14h26m01.6s and Dec +37d49m38s. The duration of the observation was 8~hours. We observed the bright source 3C\,295 as flux and bandpass calibrator. The frequency coverage for both the calibrator and cluster was 120~to~168~MHz. The visibility data was recorded every 1~second over 64~channels per 0.195~MHz sub-band. As per default, the data were flagged for interference by the observatory with AOFLAGGER \citep{offringa12}, averaged to 2~seconds and 10~channels per sub-band, and stored in the LOFAR long-term archive (LTA\footnote{\href{url}{https://lta.lofar.eu/}}).

We performed an initial direction-independent calibration following \cite{degasperin18}. Then, we corrected for direction-dependent effects using the facet calibration scheme described in \cite{vanweeren16a} and \cite{williams16}. More details are provided in the following sections.

\subsubsection{\textit{Direction-independent calibration}} 

The data were first flagged to remove the bad station CS013HBA. The bandpass and absolute flux scales were calibrated using a model of primary flux calibrator 3C\,295 from \cite{SH12}. Since the remote stations are not on the same clock as the core stations, we determined the clock offsets from the calibrator observations. After applying the calibration solutions to the target field data we performed an initial phase calibration on the target field using a model generated from the VLA Low-Frequency Sky Survey (VLSS; \citealt{cohen07}), Westerbork Northern Sky Survey (WENSS; \citealt{rengelink97}), and the NRAO/VLA Sky Survey (NVSS; \citealt{condon98}). This was followed by making high- and low-resolution images of the entire field. Compact and diffuse sources were extracted from these images with PYthon Blob Detector and Source Finder (PYBDSF: \citealt{mohanrafferty15}). These sources were then subtracted from the visibility data to create a residual (source-subtracted) data set.

\subsubsection{\textit{Direction-dependent calibration}} 

In order to improve the image fidelity, we used facet calibration to perform a direction-dependent calibration and thus correct for the ionosphere and beam errors in the direction of the cluster. As the apparent flux density of the cluster dominates the field, it was sufficient to perform a single direction-dependent calibration towards Abell~1914. The resulting calibrated data for the central facet (\amin{40} $\times$ \amin{40}) were imaged in CASA using the multi-scale multi-frequency (MS-MFS) option in the \texttt{clean()} task (\citealt{rau11}). Since the emission from Abell~1914 is within the 99\% power point of the primary beam, we did not have to correct for primary beam attenuation. While imaging, we cleaned to a depth of 2$\sigma_{rms}$ using scale sizes equal to zero, 3, 7, 9, 25, 60, and 150 pixels, where the pixels have a size of approximately one-fourth of the resolution. After cleaning, the image residuals were mostly noise-like. The brightest source in the cluster (source RP; see Figure~\ref{fig:lofar150}) has a total flux of 4.68 Jy at 150~MHz. In the presence of such a bright extended source, the local background noise of the image is negatively affected due to dynamic range limitations, most likely caused by small residual calibration errors. For this reason, two shallow negative bowls (3$\sigma$ contours) can be seen to the south-west of RP. More image details are provided in Table~\ref{tab:image}. It is important to note that the deepest negative surface brightness present in the negative region is of the order of -14$\sigma$. The RP region consists of an area of 30 beams, which means the amount of total flux density that can be suppressed because of the presence of the negative bowl is 30 $\times$ (14$\sigma$) = 0.07 Jy. With a measured flux density of 4.68 Jy for RP, this includes an extra flux of only 1.4\%. Even if, in a conservative scenario, the source is extended over 60 beams in the absence of the negative bowl, the suppression in the total flux would be 3\%, which is still small. The presence of the negative bowl may affect the radio halo flux, which is discussed later in detail. 

Owing to normalisation issues and inaccurate beam models, flux densities measured from LOFAR images need rescaling to other surveys. We used the TIFR GMRT Sky Survey (TGSS; \citealt{intema17}), which itself has a flux uncertainty of 10\%. We selected six high signal-to-noise (>20$\sigma$) compact sources within the LOFAR facet area from the TGSS to minimise the flux density measurement errors. Even though the number of sources is relatively small, the ratios of the TGSS to LOFAR flux densities are tightly clustered around 0.59 with a scatter of only 2\%. To properly take into account the propagation of all flux density uncertainties, we performed a MC simulation in which we modelled each measured LOFAR and TGSS flux density as a Gaussian distribution with the measured flux density as the centre and the uncertainty as the width of the Gaussian, and from which we determined a ensemble of flux ratios. The resulting (combined) flux ratios also follow a Gaussian distribution centred at 0.59 with a $1\sigma$ width of 0.06. We adopt these values as our flux correction factor and uncertainty, respectively. For any flux density measurement in LOFAR map, we apply the following correction:
\begin{equation}
S^t = r \times S^m
,\end{equation}
where $S^t$ and $S^m$ are the ``true'' (scaled to TGSS) and LOFAR flux densities. The value $r$ is the LOFAR flux correction factor as determined above and value $\Delta r$ is the uncertainty in the correction factor $r$. The relative error on r is $f = \Delta r / r = 0.10.$   
\begin{figure*}
\begin{center}
\resizebox{0.95\hsize}{!}{
\includegraphics[angle=0]{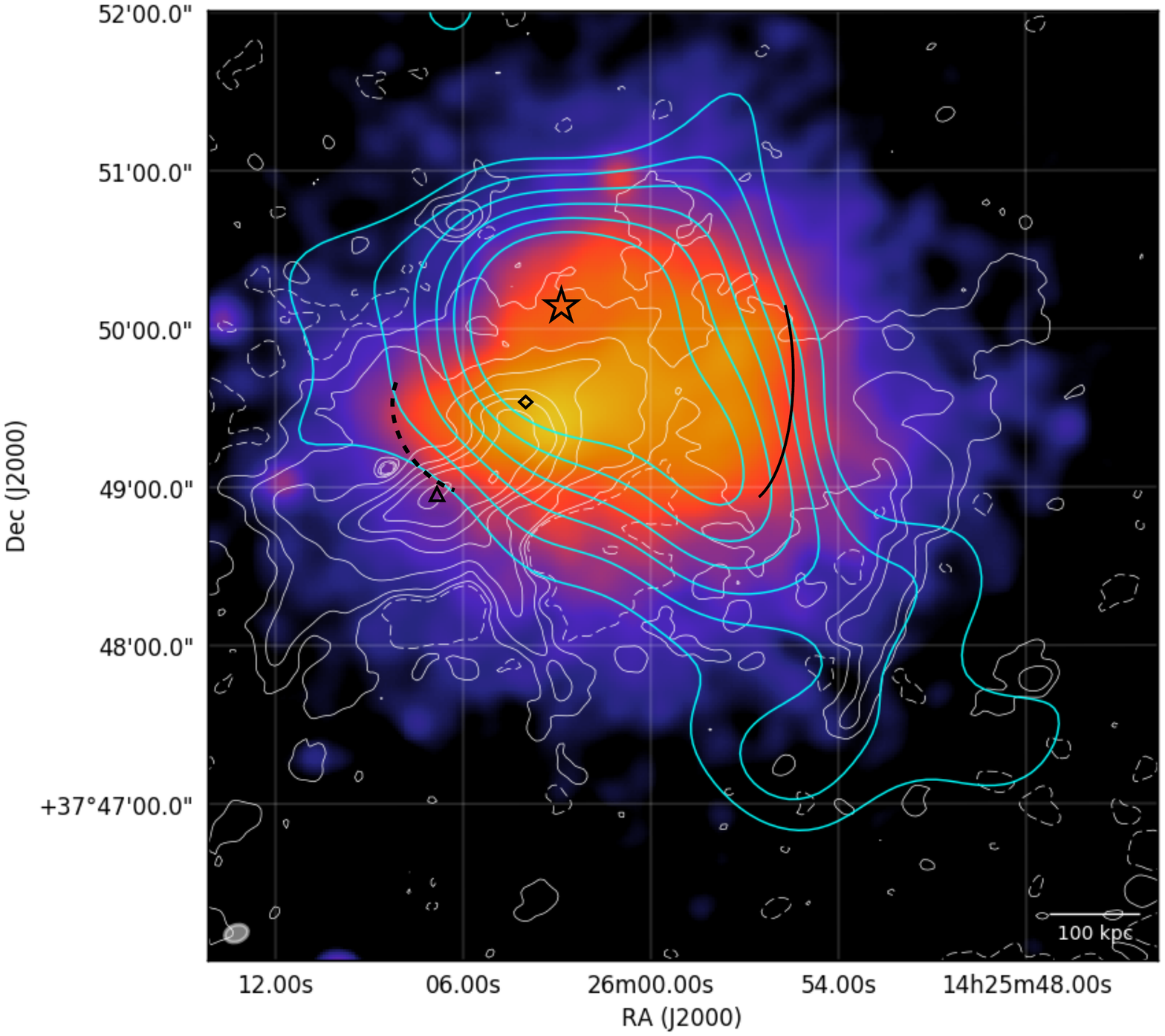}
}
\caption{Full-resolution (\asec{9} $\times$ \asec{6}) LOFAR image overlaid on an exposure-corrected, background-subtracted \textit{Chandra} image in the 0.5-2.0 keV energy band with a total integration time of 22.9ks. The LOFAR contours (white) show the (1,2,4,...)$\times$5 $\times$ $\sigma_{\rm{LOFAR}}$ levels and the dashed white contours show the -3$\sigma_{\rm{LOFAR}}$, where $\sigma_{\rm{LOFAR}}=150$~$\mu$Jy beam$^{-1}$. The cyan contours are the weak lensing measurements showing the dark matter distribution. The star, square, and triangle show the peak of the dark matter distribution, X-ray peak, and radio peak, respectively. The solid black and dashed black line show the tentative location of the shock and cold front suggested by \cite{botteon16}.}
\label{fig:xray-wl-radio}
\end{center}
\end{figure*}

\subsection{GMRT data reduction}

We used the GMRT archival data at 325~MHz of Abell~1914 (project code 16\_074) and obtained new data at 610~MHz in the GMRT observation cycle 31 (project code 31\_049; details are given in Table~\ref{tab:obs}). The data were processed using the SPAM pipeline \citep{intema17} which includes radio frequency interference (RFI) mitigation schemes, direction-dependent calibration, and ionospheric modelling \citep{intema09}. The final GMRT images at 325~MHz and 610~MHz are shown in Figure~\ref{fig:gmrt-high} and were obtained using the imaging parameters reported in Table~\ref{tab:image}. The flux density scale in the images was set by calibration on 3C48 (at 610 MHz) and 3C286 (at 325 MHz) using the models from Scaife \& Heald (2012). We adopt a flux scale uncertainty of 10\% for the GMRT observations (e.g. \citealt{chandra04}), which is quadratically added to the (random) uncertainties of all flux density measurements in the GMRT images.

\subsection{VLA data reduction}

We obtained observations of Abell~1914 at 1.4~GHz from the VLA archive (projects AF367 and AF372). The cluster was observed in C and D configurations for 160 minutes each, using two 50~MHz-wide intermediate frequency (IF) channels centred at 1.365 and 1.435~GHz; only the low-frequency IF was found to be usable in the D-configuration observation because of RFI. The data were calibrated and reduced in the standard fashion using the NRAO Astronomical Image Processing System package (AIPS). Several loops of phase self-calibration were applied to reduce the effects of residual phase errors in the data. The C- and D-configuration data sets were first processed and self-calibrated separately, then combined together to produce the final images. The flux density scale was set using the \cite{perleybutler13} coefficients, which typically results in flux errors within $5\%$. 

\subsection{Chandra data reduction}

X-ray images of Abell~1914 were produced using two ACIS-I \textit{Chandra} observations taken in \texttt{VFAINT} mode (ObsID: 542, 3593), for a total exposure time of 26.9~ks. The contamination due to soft proton flares occurring during the observations was corrected by inspecting the light curves extracted in chip~S3 using the \texttt{deflare} script. After this procedure, the net exposure time was 22.9~ks. The images we present were produced in the $0.5-2.0$~keV band adding the two ObsIDs with the \texttt{merge\_obs} routine. Data were analysed with CIAO~v4.9 and \textit{Chandra} CALDB~v4.7.3. For more detailed description see \cite{botteon18}. 

\subsection{CFHT data}

Abell~1914 was observed with MegaCam on the Canada-France-Hawaii Telescope (CFHT) as part of the Canadian Cluster Comparison Project (CCCP;\citealt{hoekstra12}, \citealt{mahdavi13}, \citealt{hoekstra15}). The main aim of this project was to study scaling relations, using weak gravitational lensing masses as reference. The $r$-band data we used were obtained with seeing $<0.8$ arcsec with a total integration time of 4800~s. Further details about the data can be found in \cite{hoekstra12}. In their most recent lensing analysis \cite{hoekstra15} determined a mass of $M_{500}=(7.3\pm 1.3)\times 10^{14} h_{70}^{-1} {\rm M}_\odot$ using the location where the X-ray emission peaks as centre. For this paper we used the lensing measurements to reconstruct the projected mass surface density using the direct inversion algorithm from \cite{kaiser93}. The approach is identical to that described in \cite{mahdavi07} in their study of the complex merger Abell~520. The reconstruction is smoothed with a Gaussian smoothing kernel with a full width at half maximum (FWHM) of $60"$, which is the consequence of the limited number density of sources in ground-based data. The resulting mass reconstruction is shown in Figure~\ref{fig:sdss-wl-radio} by purple contours. The cluster is clearly detected, as is the elongation along the merger direction. Moreover the results clearly indicate that the mass distribution peaks at the location of BCG~2.
 
\subsection{Spectral index maps and integrated spectrum calculation}

\begin{figure*}
\begin{center}
\resizebox{0.495\hsize}{!}{
\includegraphics[angle=0]{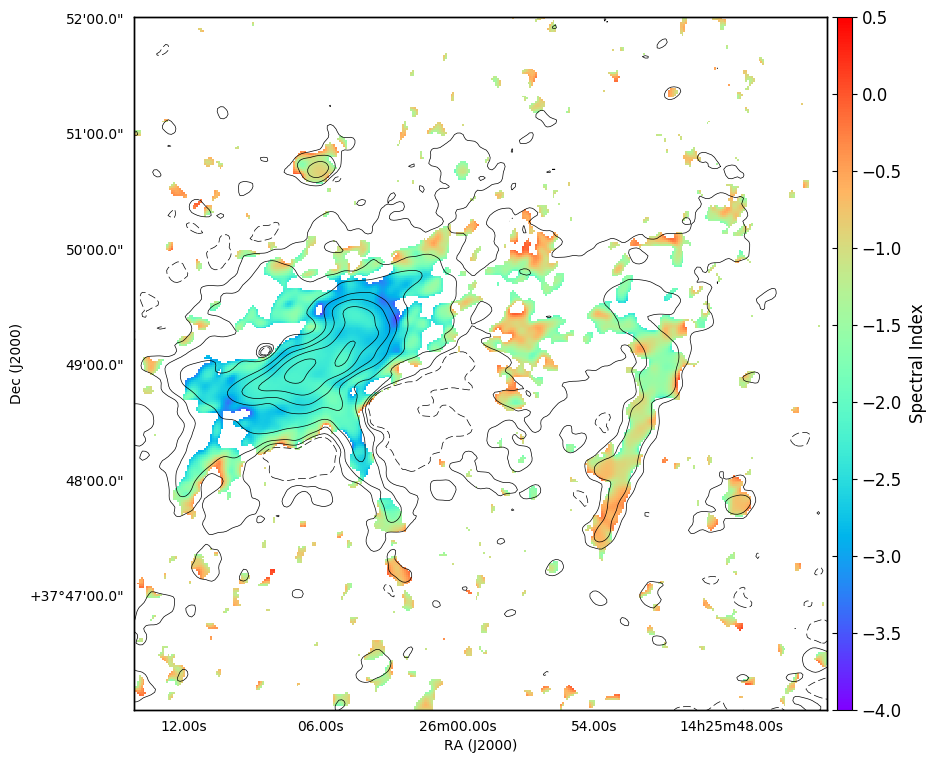}
}
\resizebox{0.495\hsize}{!}{
\includegraphics[angle=0]{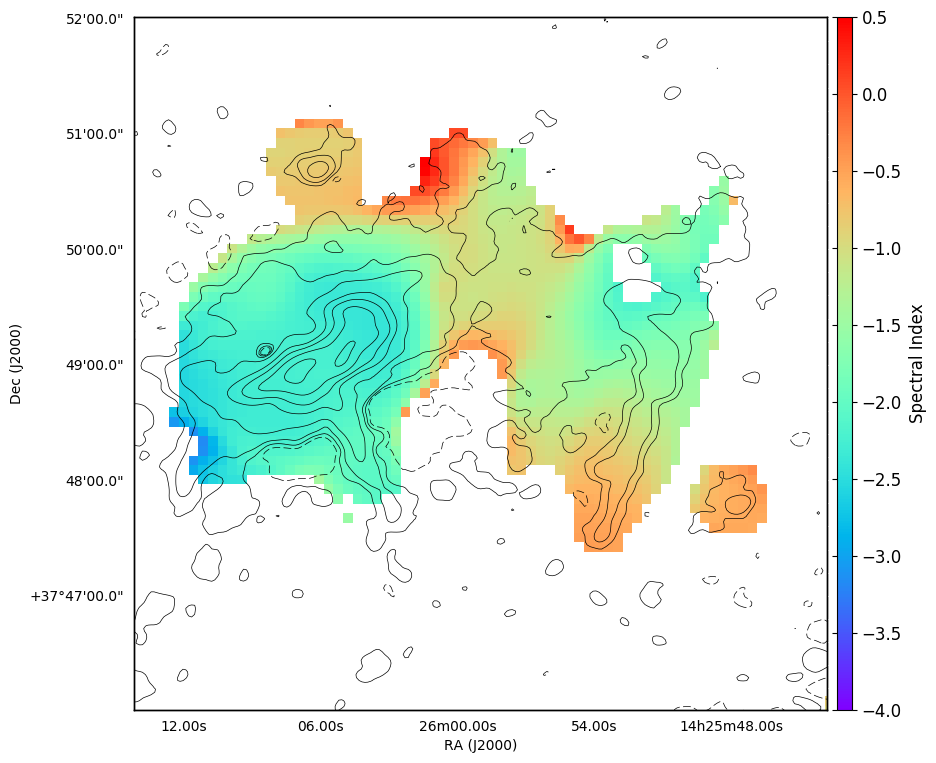}
}
\caption{Spectral properties of emission from Abell 1914 at 150~MHz to 610~MHz high-resolution (\asec{6} $\times$ \asec{6}) (\textit{left} panel) and 150~MHz--610~MHz--1.4~GHz low-resolution \asec{30} $\times$ \asec{30} (\textit{right} panel). In both figures, emission lower than $3 \sigma$ has been blanked, where $\sigma_{\rm{LOFAR_{150,6arcsec}}} = 175$~$\mu$Jy\,beam$^{-1}$, $\sigma_{\rm{GMRT_{610,6arcsec}}} = 120$~$\mu$Jy\,beam$^{-1}$, $\sigma_{\rm{LOFAR_{150,30arcsec}}} = 1500$~$\mu$Jy\,beam$^{-1}$, $\sigma_{\rm{GMRT_{610,30arcsec}}} = 230$~$\mu$Jy\,beam$^{-1}$, and $
\sigma_{\rm{VLA_{1.4,30arcsec}}} = 51$~$\mu$Jy\,beam$^{-1}$}
\label{fig:specindex}
\end{center}
\end{figure*}

Table~\ref{tab:image} lists the images made with the radio data and their corresponding resolution and sensitivity. These images were used to create a high-resolution spectral index maps between 150~MHz and 610~MHz. To sample the same spatial scales at both frequencies, we used the same inner \textit{uv} range of 200$\lambda$ to image both LOFAR and GMRT data. For the high-resolution spectral index map, we used a \textit{uv} taper of \asec{6}. The resulting images were convolved with a Gaussian to produce images with the same restoring beam, accurately aligned in the image plane, and regridded onto the same pixel grid. Surface brightness below 3$\sigma$ in respective images were masked. These images were then used to create spectral index maps where a power-law spectral index was calculated for each pixel. The error on the fit was calculated given the error on the LOFAR and GMRT pixel values ($\Delta S_1$ and $\Delta S_2$)   with
\begin{equation}
\Delta \alpha = \frac{1}{ \ln \frac{\nu_1}{\nu_2}} \sqrt{\big(\frac{\Delta S_1}{S_1}\big)^{2} + \big(\frac{\Delta S_2}{S_2}\big)^{2}}
,\end{equation}
where $\Delta S_{i} = \sqrt{(\sigma_i)^2 + (fS_i)^2}$ (with $i=[1,2]$) is the total uncertainty associated with the measured flux density $S_i$ and the subscripts stand for 150 MHz and 610 MHz, respectively, and $f$ is the flux scale uncertainty as described in Section 2.1.2, for the LOFAR image. The total error propagation takes into account the flux scale uncertainty and, $\sigma_i$, the rms noise of the respective image. A similar procedure was used to create a low-resolution spectral index map between 150~MHz and 1.4~GHz. The resulting spectral index images are shown in Figure \ref{fig:specindex}.

\section{Results}

In Figure~\ref{fig:lofar150} we labelled three main regions of diffuse emission associated with Abell~1914 which are discussed in the following sections.

\subsection{RP: A radio phoenix candidate}

The source RP is arguably the most interesting feature of this system. The source, classified as 4C38.39 is known to have a steep spectrum ($\alpha \sim$ -2) and a possible curvature (\citealt{kg94}, \citealt{roland85}). The RP consists of a bright, resolved region surrounded by fainter diffuse emission (see Figure~\ref{fig:lofar150}). The bright region extends $\sim 650$~kpc in the NW-SE direction and $\sim 230$~kpc in the perpendicular direction. In the GMRT 325~MHz and 610~MHz images (Figure~\ref{fig:gmrt-high}) the source RP is clearly detected, but the diffuse component surrounding it is marginally detected at 325~MHz. Figure~\ref{fig:lofar-vla} shows the low-resolution 150~MHz LOFAR and VLA 1.4~GHz images. 

\begin{figure}
\begin{center}
\resizebox{1.0\hsize}{!}{
\includegraphics[angle=0]{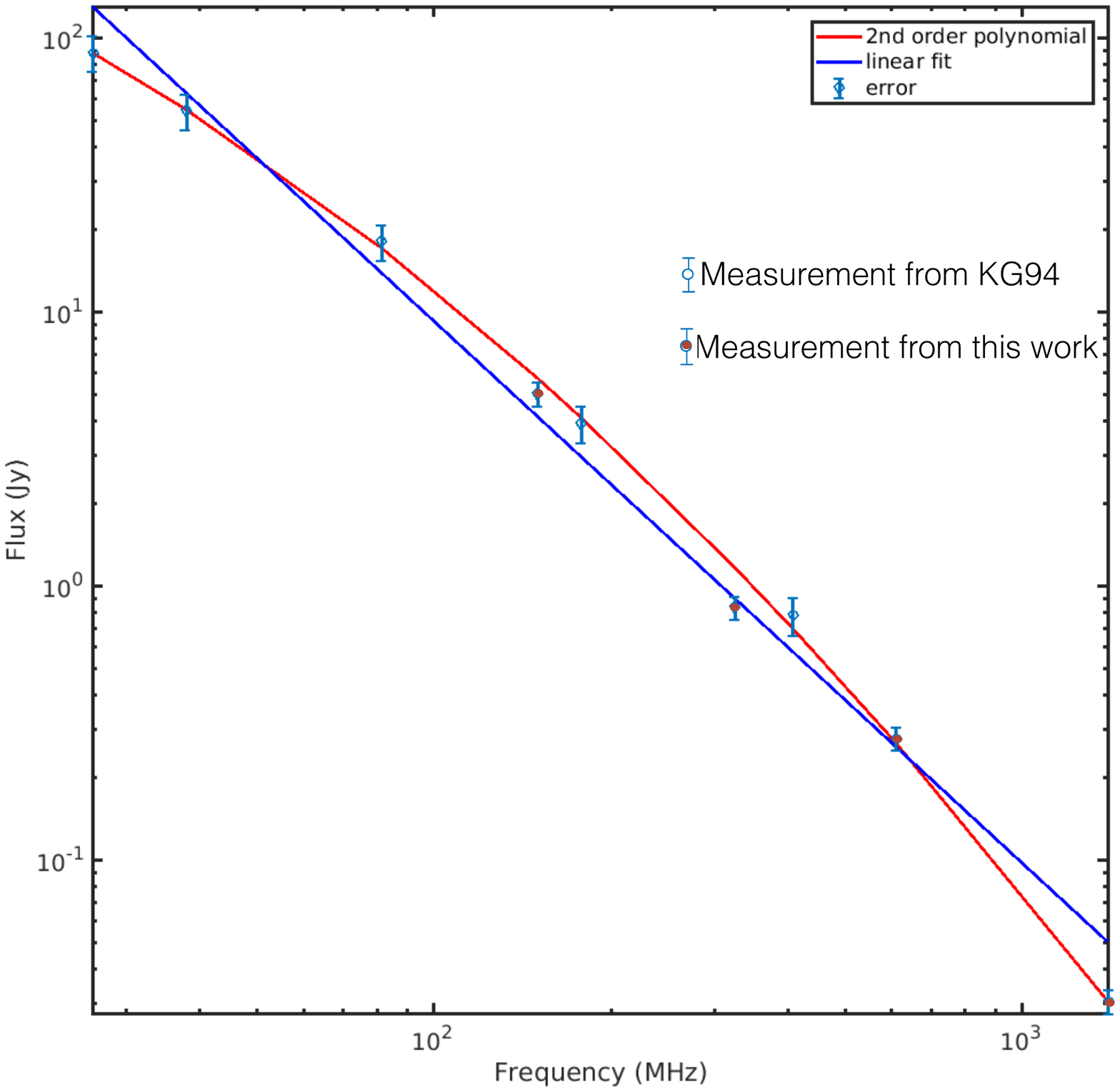}
}
\caption{Total flux density as a function of frequency of the source RP. The brown points show the flux measurements at 150~MHz, 325~MHz, 610~MHz, and 1.4~GHz (this paper), while the blue points show the measurements from \cite{kg94}. 
}
\label{fig:spectra}
\end{center}
\end{figure}

In order to measure the integrated spectral index of the source RP we made a set of new images (not shown) at all available radio frequencies with the same imaging parameters and matched minimum \textit{uv} range. We measured the flux density over exactly the same area, defined by the $3\sigma$ contour of the LOFAR image (Figure~\ref{fig:lofar150}) while excluding the RH region. The measured total flux densities at 150 MHz, 325 MHz, 610 MHz, and 1.4 GHz are $\rm{S_{150}}$ = 4.68$\pm$0.46 Jy, $\rm{S_{325}}$ = 0.83$\pm$0.08 Jy, $\rm{S_{610}}$ = 0.277$\pm$0.02 Jy, and $\rm{S_{1.4}}$ = 34.8$\pm$2.0 mJy, respectively. A single power-law fit gives an integrated spectral index measurement of $\alpha_{150-1400} = -2.17 \pm 0.11$ for the source RP. We used the LOFAR 150~MHz and GMRT 610~MHz higher resolution maps to create a resolved spectral index map. Figure~\ref{fig:specindex} reveals a fairly uniform spectral index distribution with small variations; the spectral index value ranges from -1.9 to -2.3. The spectral index tends to get flatter in the southern direction. We used the additional flux measurements from \cite{kg94} and plotted the spectrum of RP (Figure~\ref{fig:spectra}). We note that measurements in \cite{kg94} were obtained using different imaging parameters than ours, which may have an impact on the relative accuracy between their measurements and ours. We fitted a line and a second order polynomial in log-log space to the data. The goodness of the fits was determined by the parameter sum of squares due to error (\textit{SSE}), which measures the total deviation of the response values from the fit to the response values. The \textit{SSE} values are 0.70 and 0.13 for the linear and second order polynomial fit, which suggests that the second order polynomial fit the data points better than the linear fit. Therefore, the spectrum shows hints of possible spectral curvature, as suggested in \cite{kg94}. In order to confirm the curvature, we need additional, accurate measurements, especially at the lowest frequencies such as future LOFAR LBA (Low Band Antenna) observations.

In Figure~\ref{fig:sdss-wl-radio} the weak lensing mass reconstruction based on the CFHT optical image shows that the total mass distribution (dominated by dark matter) roughly traces the galaxy distribution. The peak of the radio emission of RP  (triange marker) does not coincide with the nearby BCG positions (star markers). RP has an interesting extension towards the south-west, which we associate with an optical counterpart (circle marker, 
labelled RG). While an apparent tail of radio emission extends from RG towards RP, it is unclear if all radio emission in RP originates from RG. We do not identify any other obvious optical counterparts for RP. 
In Figure~\ref{fig:xray-wl-radio} the peak of the radio emission (triangle marker) does not coincide with the X-ray peak (square marker). The X-ray emission is elongated in the SE-NW direction, which is a similar orientation to the weak lensing mass reconstruction, but the latter is more elongated. 

As a consequence of the extreme steep spectrum, combined with the filamentary morphology, peripheral location, misalignment with X-ray, and weak-lensing mass distribution, the source RP does not match the typical observational properties of giant radio haloes or radio relics, but it could be a radio phoenix. This is discussed in more detail in Section~\ref{sec:discussion}.

\begin{figure*}
\includegraphics[scale=0.6]{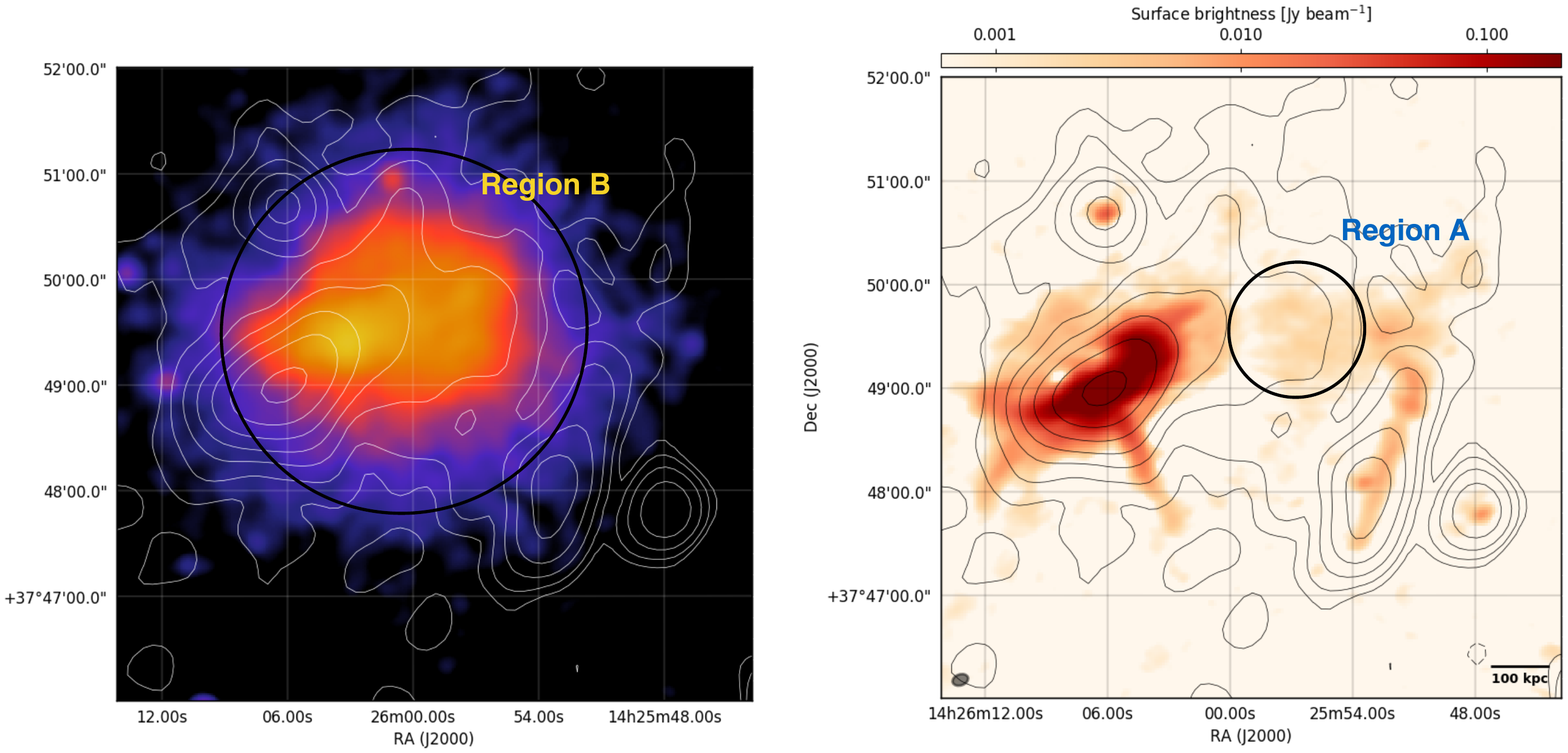}
\caption{\textit{Left panel:} Low-resolution, \asec{30} $\times$ \asec{30} VLA 1.4~GHz image contours overlaid on the \textit{Chandra} image. The VLA contours (white) show the (1,2,4,...)$\times$3$\times$ $\sigma_{\rm{VLA}}$, where $\sigma_{\rm{VLA}}$ = 55~$\mu$Jy\,beam$^{-1}$. Region B was used to extrapolate the radio halo flux (see section 3.3). \textit{right panel:} The same low-resolution VLA image contours (in black) overlaid on the LOFAR 150~MHz extended emission subtracted image. Region A shows the portion where the radio halo emission is not contaminated by the RP and HT.}
\label{fig:vla-lofar-cor}
\end{figure*}

\subsection{HT: A head-tail galaxy}

We identify the source HT as a head-tail galaxy. In Figure~\ref{fig:sdss-wl-radio} the CFHT $r$-band optical image is overlaid with the LOFAR 150~MHz map contours shows the optical counterpart of the head of the HT (marked as A). Another optical counterpart (B) is not a member of Abell~1914 \citep{sifon15}. The extent of the tail is 630~kpc in the NW direction. From the high-resolution spectral index map (Figure~\ref{fig:specindex} left panel) we observe the typical steepening of the radio spectrum (-0.89 $\geq \alpha \geq$ -1.15) along the length of the tail. The emission towards the end of the tail blends into the diffuse emission around the region of RP, which we discuss in the following section.

\subsection{RH: A radio halo candidate}
\label{sec:rhc}

The faint emission region between sources RP and HT in Figure~\ref{fig:lofar150}, labelled with RH, appears to be morphologically disconnected from the neighbouring sources. The region RH resembles radio halo emission, but cannot be traced over the full extend of the X-ray emission owing to the contaminating presence of both RP and HT. Large-scale emission in this region is also seen in low-resolution LOFAR 150 MHz and VLA 1.4 GHz images (Figure \ref{fig:lofar-vla}). Assuming it is part of a radio halo, we selected the region~A ($\sim 70$~kpc) as shown in Figure~\ref{fig:vla-lofar-cor} (\textit{right panel)} to calculate the flux density and spectral index. In this region, the halo emission is likely separated from the emission from RP and HT. From the 150~MHz LOFAR and 1.4~GHz VLA images we measure the integrated spectral index of the halo in region~A to be $\alpha$ = -1.15 $\pm 0.20$.

We estimate the radio halo power at 1.4~GHz using the VLA image. The measured total flux density of the region A is 2.34 $\pm$ 0.15 mJy. Using the 1.4~GHz flux measurements we calculated a \textit{k}-corrected 1.4~GHz radio power $P_{\rm{1.4GHz}}$ = $4\pi S_{\rm{1.4GHz}}D^2_{L}(1+z)^{-\alpha-1} = (1.94 \pm 0.14) \times 10^{23}$~W\,Hz$^{-1}$. The value of the radio power at 1.4~GHz is underluminous when compared with the $P_{1.4} - M_{500}$ relation presented in \citealt{cassano13}. However, because of the contamination from RP, HT, and negative bowls around RP, the radio halo flux may be significantly underestimated. To show an example, we assume that the radio halo spans the region~B ($\sim$ 1~Mpc diameter) as shown in Figure~\ref{fig:vla-lofar-cor} (\textit{left panel}). To calculate the total radio halo flux density in region~B, we assume that it has the same surface brightness as region~A. The extrapolated total flux density of region~B is $18.72\pm1.23$~mJy. We note that the error does not properly reflect the intrinsic uncertainty of our assumptions. With this value of the radio halo flux density, we measure a \textit{k}-corrected 1.4~GHz radio power $P_{\rm{1.4GHz}} = (1.55 \pm 0.11) \times 10^{24}$~W\,Hz$^{-1}$. This value of the radio power falls within the scatter of the correlation presented in \citealt{cassano13}.

We also attempted to separate possible radio halo emission from the emission of sources RP and HT by re-imaging the LOFAR data while leaving out the shortest baselines, which suppresses the extended emission and leaves only compact source emission. In principle, this source model could then be subtracted from the visibility data, leaving only extended emission in the data. However, the clean subtraction of a complex, bright source such as RP on a scale of $> 100$~kpc turns out to be too inaccurate to be of much practical use. 

\section{Discussion}
\label{sec:discussion}

The radio emission in Abell~1914 has previously been classified as an irregularly shaped radio halo with a steep spectrum (B03, G04) and a hint of spectral curvature (\citealt{kg94}). With the new LOFAR observations, combined with archival GMRT, VLA, CFHT, and \textit{Chandra} data, we are able to separate the emission into three different components: a candidate radio phoenix, a possible radio halo and a tailed radio galaxy. These components are discussed below.

\subsection*{Radio phoenix}

\cite{kg94} suggested that the USS radio source 4C38.39 in Abell~1914 may indicate the presence of a very old radio galaxy. When the core of the radio galaxy had ceased its activity, the supply of fresh relativistic electrons into the radio lobes also stopped. Synchrotron losses of these electrons can explain the unusually steep spectrum nature. However, more recent studies of this source suggested it was part of a radio halo (B03, G04). The LOFAR and GMRT observations show that the properties of the emission in this region are quite distinct from those expected for a radio halo. Instead we note that this source very much resembles a radio phoenix source because (i) it has an unusually steep spectrum, (ii) the morphology of the source is quite filamentary when compared to a classical radio halo morphology, (iii) the X-ray morphology is displaced from the radio emission, and (iv) the source shows hints of having a curved spectrum.

Radio phoenixes are aged radio galaxy lobes whose emission is boosted by adiabatic compression (e.g. \citealt{eg01}, \citealt{ensslinbruggen02}). There is evidence of these kind of sources in literature (e.g. \citealt{slee01}, \citealt{vanweeren09}, \citealt{vanweeren11}, \citealt{ogrean11}, \citealt{cohen11}, \citealt{stroe13}, \citealt{gasperin15}). There is an interesting connection between the source RP and the radio galaxy (indicated with RG in Figure \ref{fig:sdss-wl-radio}). RP can be a radio bubble detached by a shock from the tail of the radio galaxy (e.g. \citealt{jones17}) unless the connection is just a projection effect. Besides adiabatic compression, shocks that are introduced by merging events might also be able to re-accelerate electrons via Fermi-type processes. However the hint of curvature and the fact that the source is very steep may rule out the possibility of shock acceleration, since in the case of shock acceleration the spectrum tends to get flatter (e.g. \citealt{bonafede14}, \citealt{botteon16}, \citealt{vanweeren17}). Instead the above facts are more consistent with the adiabatic compression scenario. Therefore, detection of shocks near the location of RP is crucial to confirm the radio phoenix compression scenario. 

\cite{botteon18} suggested that the merging scenario in Abell~1914 is similar to that observed in the bullet cluster \citep{markevitch02} where a subcluster is moving from the E to the W direction, producing a cold front in the direction of the motion. On the eastern side of Abell~1914, they claimed the presence of a shock moving in the cluster outskirts, similar to the reverse shock found in the Buller cluster \citep{shimwell15}. Deeper X-ray observations are needed to confirm this.

\subsection*{Radio halo}

In previous work, Abell~1914 was suggested to host a steep spectrum radio halo (B03, G04), but there the bright source 4C38.39 was assumed to be part of the halo. While the latter is likely not true, we do detect diffuse radio emission in between 4C38.39 and the tailed radio galaxy that might be part of a larger radio halo. The spectral index of the region RH is flatter than the tail of HT and RP (Figure \ref{fig:specindex} \textit{right} panel). The measured, \textit{k}-corrected 1.4~GHz radio power of the radio halo (region A in Figure \ref{fig:vla-lofar-cor}) in Abell~1914 (discussed in Section~\ref{sec:rhc}) is $\rm{P_{\rm{1.4GHz}}} = (1.94 \pm 0.14) \times 10^{23}$ W/Hz. It is known that the radio power at 1.4~GHz and mass of the hosting cluster are correlated for giant radio haloes (e.g. \citealt{cassano13}). According to our estimates (Section \ref{sec:rhc}) the luminosity of the halo in Abell~1914 is significantly smaller than previously thought (\citealt{cassano13}) and the halo appears to be  underluminous with respect to expectations based on the radio power -- mass correlation. Predictions suggest that USS radio haloes are underluminous when compared with the correlation \citep{cassano10a}. However, at least for the region A, the low-resolution spectral index map between 150~MHz LOFAR, 325~MHz GMRT, 610~MHz GMRT, and 1.4~GHz VLA (Figure~\ref{fig:specindex}; \textit{right panel}) suggests a spectral index of $\alpha = -1.15 \pm 0.20$. With the availability of current data there is no evidence of the radio halo being ultra-steep. 

Recently hadronic models have also been invoked to explain underluminous radio haloes \citep{cuciti18}, but in these cases the radio halo should be centred on the X-ray peak location. {So these models cannot explain the scenario in which the radio halo in Abell 1914 is underluminous. However, it is important to note that the measurement of the radio halo power is highly uncertain owing to the contamination from RP and HT. In Section \ref{sec:rhc} we also estimated the radio halo flux and corresponding radio halo power in a much larger region B (Figure \ref{fig:vla-lofar-cor}) indicating that the value of the radio power can lie in between these two measurements. Higher resolution observations at GHz frequencies are needed to confirm the spectral properties of the radio halo.   

\subsection*{Head-tail galaxy}

The head of the head-tail galaxy on the west side of Abell~1914 was reported as a discrete source by B03, while the diffuse tail was unresolved and blended together with the other diffuse sources. In this paper, with the help of detailed spectral index maps, we classify this source unambiguously as a head-tail radio galaxy. From the spectral index maps (Figure~\ref{fig:specindex}) a clear spectral steepening can be observed towards its tail due to the synchrotron ageing. The disrupted X-ray morphology, mass distribution, and presence of a radio halo indicates that the cluster is undergoing a merger, but its effect on the HT is not clear. Initially the tail is more collimated and undisturbed, whereas the end of the tail seems to be more disrupted and blended with the radio halo emission. The spectral index at the end of the tail is similar to that of the radio halo emission, which could be the evidence for re-acceleration. In this scenario, old electrons in the turbulent end of the tail are mixed with the ICM and re-accelerated in the same way as radio halo particles are accelerated (which is still debated). \cite{botteon18} suggested a presence of a shock (Figure~\ref{fig:xray-wl-radio}) near the location in between RH and HT, which could also play a role in local particle re-acceleration.

\section{Summary and conclusions}

In this paper, we presented new LOFAR HBA radio observations at 150~MHz of the merging galaxy cluster Abell~1914. We also analysed GMRT and VLA data and presented high- and low-resolution spectral index maps. These allowed us to constrain spectral properties of this system. We summarise the main results as follows:

Abell~1914 was thought to host a radio halo but with quite unusual properties. With the new LOFAR observations at high resolution (\asec{9} $\times$ \asec{6}) we were able to properly resolve the radio structure, which was not possible in previous studies. With the new resolved radio maps, we find that the overall radio emission of Abell~1914 is unlikely to be a single radio halo but rather the superposition of a revived fossil plasma source, a radio halo, and a head-tail radio galaxy.

We characterise the revived fossil plasma source as a radio phoenix candidate. The source RP (Figure~\ref{fig:lofar150}) has an integrated spectral index measurement of $-2.17 \pm 0.11$ with a possible curvature (Figure~\ref{fig:spectra}). Usually these sources are thought to be old fossil plasma, originating from a dead radio galaxy, which has been adiabatically compressed by the passage of shocks (\citealt{eg01}; \citealt{ensslinbruggen02}). The hint of curvature and ultra-steep nature probably rules out the possibility of shock acceleration. \cite{botteon18} do not find clear evidence of a shock near the vicinity with the present shallow X-ray observations. Deeper X-ray observations are needed to confirm this.

The possible presence of a faint radio halo is inferred from emission in a region that seems not to be contaminated by radio phoenix or head-tail emission. Based on our estimates, the radio halo appears underluminous with respect to its mass, as predicted by the $P_{1.4GHz}-M_{500}$ correlation plot presented by \cite{cassano13}. Usually USS radio haloes are underluminous in the correlation plot. However, in the region of the halo where it is possible to measure the spectrum, we find the spectral index to be $-1.15 \pm 0.20$. It is important to note that the measured value of the radio power is uncertain and may significantly be underestimated because of the possible contamination from RP and HT.

From the radio intensity and spectral index maps (Figures~\ref{fig:lofar150} and \ref{fig:specindex}) we identify a head-tail radio galaxy with an optical counterpart in the head and a clear spectral steepening towards its tail. One interesting aspect is the connection of the tail with the assumed radio halo emission, which could be a source of mildly relativistic electron injection. A few other examples of possible connections between radio haloes and head-tail galaxies are found in \cite{wilber18} and \cite{rajpurohit18}.

\section*{Acknowledgments}
We thank the anonymous referee for useful comments. This paper is based (in part) on data obtained with the International LOFAR Telescope (ILT) under project code LC0\_037. LOFAR \citealt{vanhaarlem13} is the LOw Frequency ARray designed and constructed by ASTRON. It has observing, data processing, and data storage facilities in several countries, which are owned by various parties (each with their own funding sources) and are collectively operated by the ILT foundation under a joint scientific policy. The ILT resources have benefited from the following recent major funding sources: CNRS-INSU, Observatoire de Paris and Université d'Orléans, France; BMBF, MIWF-NRW, MPG, Germany; Science Foundation Ireland (SFI), Department of Business, Enterprise and Innovation (DBEI), Ireland; NWO, The Netherlands; and The Science and Technology Facilities Council, UK. We thanks the staff of the GMRT who made these observations possible. GMRT is run by the National Centre for Radio Astrophysics of the Tata Institute of Fundamental Research. This paper is based on the data obtained with the International LOFAR Telescope (ILT). RJvW acknowledges support from the ERC Advanced Investigator programme NewClusters 321271 and the VIDI research programme with project number 639.042.729, which is financed by the Netherlands Organisation for Scientific Research (NWO). DNH acknowledges support from the ERC Advanced Investigator programme NewClusters 321271. F.d.G. is supported by the VENI research programme with project number 1808, which is financed by the Netherlands Organisation for Scientific Research (NWO). AD acknowledges support by the BMBF Verbundforschung under the grant 05A17STA. Basic research in radio astronomy at the Naval Research Laboratory is supported by 6.1 Base funding. This research made use of APLpy, an open-source plotting package for Python hosted at \href{url}{http://aplpy.github.com.}

\bibliographystyle{aa} 
\bibliography{paper.txt} 

\label{lastpage}
\end{document}